\documentclass[12pt]{article}
\usepackage[cm]{fullpage}
\usepackage{amsmath}
\usepackage{amsfonts}
\usepackage{amssymb}
\usepackage{amsthm}
\usepackage{graphicx}
\usepackage{fancyhdr}
\usepackage{xcolor}

\usepackage{wasysym}

\definecolor{bb}{rgb}{0.3, 0.5, 1}
\definecolor{bg}{rgb}{0.1, 0.1, 0.5}
\usepackage[%
    pdftitle={The title of your letter},
    pdfauthor={Your name},
    pdfsubject={The subject of the letter},
    pdfkeywords={Any keywords},
     colorlinks=true,  
     linkcolor=bg, 
     citecolor=bg, 
     urlcolor=bg, 
     pdfnewwindow=true,
     pagebackref=true,
     filecolor=bb,
     pdffitwindow=false,     
     pdfstartview={FitH}    
]{hyperref}

\def\ba{\begin{eqnarray}}
\def\ea{\end{eqnarray}}
\def\be{\begin{equation}}
\def\ee{\end{equation}}

\def\d{\mathrm{d}}

\def\mn{_{\mu \nu}}

\def\({\left(}
\def\){\right)}

\def\mpl{M_{\rm Pl}}

\begin{document}

\hspace{4.8in} \mbox{NORDITA-2015-62}\\\vspace{-1.03cm} 

\vskip 1.6cm

\centerline{\Large \bf Non-minimal derivative couplings of the composite metric}
\vskip 0.7cm
\centerline{\large Lavinia Heisenberg\footnote{laviniah@kth.se}}
\vskip 0.3cm

\centerline{\em Nordita, KTH Royal Institute of Technology and Stockholm University, }
\centerline{\em Roslagstullsbacken 23, 10691 Stockholm, Sweden }
\centerline{\em Department of Physics \& The Oskar Klein Centre, }
\centerline{\em AlbaNova University Centre, 10691 Stockholm, Sweden}

\vskip 1.9cm

\begin{abstract}

In the context of massive gravity, bi-gravity and multi-gravity non-minimal matter couplings
via a specific composite effective metric were investigated recently. Even if these couplings
generically reintroduce the Boulware-Deser ghost, this composite metric is unique in the sense
that the ghost reemerges only beyond the decoupling limit and the matter quantum loop corrections
do not detune the potential interactions. We consider non-minimal {\it derivative} couplings of the
composite metric to matter fields for a specific subclass of Horndeski scalar-tensor interactions. We first
explore these couplings in the mini-superspace and investigate in which scenario the ghost remains absent.
We further study these non-minimal derivative couplings in the decoupling-limit of the theory and show
that the equation of motion for the helicity-0 mode remains second order in derivatives. Finally, we discuss 
preliminary implications for cosmology.

\end{abstract}

\vspace{1cm}

\section{Introduction}

The accelerated expansion of the late Universe is confirmed by many cosmological observations like supernovae, CMB, baryon acoustic oscillations and lensing. It is believed that this expansion is driven by dark energy, whose origin has not yet been identified. There are many attempts to explain the nature of dark energy. Assuming that General Relativity is the ultimate theory for gravitational interactions, one can relate the effective repulsive force between cosmological objects with the presence of a cosmological constant $\Lambda$ with a constant energy density. In fact, from an effective field theory point of view it must be included. From a particle physics perspective the origin of the cosmological constant could be related to the vacuum energy density. Nevertheless, the theoretical expectations for it using standard techniques of quantum field theory exceeds the observational bounds by a tremendous amount. This constitutes the century problem of modern cosmology \cite{Weinberg:1988cp, Straumann:2002tv}. 

As an alternative explanation for the recent acceleration of the Universe one can invoke new dynamical degrees of freedom, either in form of new fluids with negative pressure in the stress energy tensor, or in form of modifications of the geometrical part of Einstein's equations. The main goal of these approaches is to weaken gravity on cosmological scales. These new degrees of freedom could play the role of a condensate, whose energy density either sources self-acceleration or compensates the cosmological constant. Such scenarios could for instance arise in massive gravity or in higher-dimensional frameworks.

Concerning the higher dimensional framework, the Dvali-Gabadadze-Porrati (DGP) model was a promising candidate among large scale modified theories of gravity \cite{Dvali:2000hr}. In this scenario our Universe is confined to a three-brane embedded in a five-dimensional bulk.
The brane curvature in form of an intrinsic Einstein Hilbert term guarantees the recovery of General Relativity on small scales. On the other hand, on cosmological scales the modifications can be interpreted as a soft mass of the graviton, that systematically limits the effective range of the graviton. A massless higher dimensional graviton carries the same number of degrees of freedom as a massive graviton on four dimensions, hence the effective graviton on the brane carries the usual helicity-2 modes, two helicity-1 modes and one helicity-0 mode as in massive gravity. The phenomenological consequences of these theories is tightly related to the helicity-0 mode, which can mediate an extra fifth force. The helicity-1 modes usually decouple. Even in the presence of this helicity-0 mode one can recover General Relativity via the Vainshtein mechanism \cite{Vainshtein:1972sx}. It relies on the successful implementation of screening the unwanted modes from the gravitational dynamics on small scales via nonlinear interactions. The coupling of the helicity-0 mode to matter is suppressed once one canonically normalises the fields. This decoupling of the nonlinear helicity-0 mode is manifest in the so-called decoupling limit. Thanks to this helicity-0 mode, the DGP model admits self-accelerating solutions, which however are plagued by ghost-like instabilities \cite{Deffayet:2001pu, Koyama:2005tx}. This gave rise to attempts for considering more exhaustive setups with Gauss-Bonnet terms in the bulk \cite{deRham:2006pe}.

The decoupling limit of DGP contains a very specific derivative interactions for the helicity-0 mode. Their form is dictated by the residuals of the 5-dimensional Poincar\'e invariance. Soon it was realised that the decoupling limit of DGP can be extended to more general interactions with invariance under internal Galilean and shift transformations \cite{Nicolis:2008in}. Together with the postulate of second order equations of motion these symmetries severely restrict the allowed derivative interactions. These Galilean invariant interactions have been treated in their own right as scalar field theories and received much attention for their rich phenomenology \cite{Silva:2009km, Chow:2009fm, Creminelli:2010ba, Burrage:2010cu, Wyman:2011mp}. In the presence of a compact source these Galileon theories admit spherically symmetric solutions that propagate superluminal \cite{Hinterbichler:2009kq, Nicolis:2008in, deFromont:2013iwa}, even though a direct analogue of Hawking's chronology protection conjecture prevents to construct closed timelike curves \cite{Burrage:2011cr}. The generalisation of the Galileon to a non-flat background breaks the Galilean symmetry. It was also realised that one has to invoke non-minimal coupling between the scalar field and the curvature in order to maintain the property of second order equations of motion \cite{Deffayet:2009wt}. The same covariant interactions can be also obtained from a higher-dimensional construction \cite{deRham:2010eu}. Furthermore, the Galilean symmetry can be promoted to a generalized Galileon symmetry on maximally symmetric backgrounds. For instance, the de Sitter Galileon interactions acquire additional, specific potential-like terms \cite{Goon:2011qf, Burrage:2011bt}.

Interestingly, these Galilean invariant interactions naturally arise in gravitational theories with a massive graviton as an exchange particle \cite{deRham10, deRham:2010tw}. The existence of a graviton mass is a fundamental question from a theoretical perspective. It was a challenge to construct a consistent non linear theory for massive gravity without the Boulware-Deser ghost, which has witnessed a tremendous amount of progress in the past decade \cite{deRham10, dRGT10, deRham11, Hassan:2011vm, Hassan:2011hr, Hassan:2012qv}, and shown to be technically natural \cite{deRham:2012ew, deRham:2013qqa}. Besides the interesting analogue to Galileon interactions, one can also connect massive gravity to Horndeski type of theories. Horndeski interactions are the most general scalar-tensor interactions which give rise to second order equations of motion \cite{Horndeski, Deffayet:2009mn}. The covariantization of the decoupling limit of massive gravity yields a proxy theory to massive gravity \cite{deRham:2011by}, which is a non-minimally coupled subclass of Horndeski scalar-tensor theory, which contains four arbitrary functions in the Lagrangian. In the proxy theory, these four arbitrary functions are dictated by the covariantization procedure, and it shares the same decoupling limit with massive gravity. The phenomenological implications of this proxy theory were studied in \cite{deRham:2011by, Heisenberg:2014kea}. 

The Galileon interactions are non-renormalizable. Exactly the same property is also fulfilled by the decoupling limit interactions of massive gravity \cite{deRham:2012ew, Heisenberg:2014raa}. Of course this is not the case anymore when one considers also matter loop quantum corrections. In this context, it is crucial to investigate the possible consistent couplings to matter fields in massive gravity. One important requirement is to couple the matter fields in a way that does not detune the potential interactions at the quantum level. There is an unique composite effective metric built out of the two metrics that fulfils this requirement and guarantees the absence of the Boulware-Deser ghost in the decoupling limit \cite{deRham:2014naa, deRham:2014fha, Noller:2014sta}. The presence of this effective composite metric has important theoretical and phenomenological consequences \cite{deRham:2014naa, Akrami:2013ffa, Akrami:2014lja, Hassan:2014gta, Heisenberg:2014rka, Enander:2014xga, Schmidt-May:2014xla, Mukohyama:2014rca, Solomon:2014iwa, Gao:2014xaa, Comelli:2015pua, Gumrukcuoglu:2014xba, Gumrukcuoglu:2015nua, Hinterbichler:2015yaa, deRham:2015cha, Huang:2015yga, Heisenberg:2015iqa, Blanchet:2015sra, Blanchet:2015bia}. In this work we would like to further study the theoretical implications of this effective composite metric in the context of non-minimally coupled scalar-tensor theories motivated by the proxy theory. For this purpose, we consider the specific scalar-tensor interactions of the covariantized decoupling limit, that live on the effective composite metric. It is important to investigate whether non-minimal couplings to this effective metric maintain the ghost absent in the decoupling limit, and whether this might yield richer cosmological phenomenology than the standard Horndeski type of theories living in the standard metric.


\section{Non-minimal derivative couplings}\label{sec:MG1}
An interesting question within massive (bi-) gravity that has received much attention lately is the consistent coupling to matter fields. In this context a new composite effective metric was proposed in \cite{deRham:2014naa} and it was shown, that it is the unique metric that removes the Boulware-Deser ghost in the decoupling limit and through which the matter loop quantum corrections do not detune the potential interactions. For an application to dark matter, in \cite{Blanchet:2015sra} it was shown that the kinetic Lagrangian needs to be restricted to either $M_g^2 \sqrt{-g} R_g + M_f^2\sqrt{-f} R_{f}$ as in the standard ghost-free kinetic terms in bigravity with dynamical $g$ and $f$ metrics \cite{Hassan:2011zd}, or alternatively one can replace one of these kinetic terms by the corresponding kinetic term of this new composite effective metric $M_g^2\sqrt{-g} R_g + M_\text{eff}^2\sqrt{-g_\text{eff}} R_\text{eff}$. It was proven that one has to restrict the theory to have not more than two kinetic terms in order not to reintroduce the BD ghost in the decoupling limit. In this work, we go along the line of investigations as in \cite{Blanchet:2015sra}, but aim to push forward this analysis to non-minimal derivative couplings to matter fields and investigate whether these derivative couplings reintroduce the ghost in the decoupling limit and weather the ghost propagates on top of maximally symmetric backgrounds. For this purpose we choose to work with the following action

\begin{equation} \label{eq:TotalAction}
S=\int\sqrt{-g}\mpl^2  R + m^2M_{\text{eff}}^2\sqrt{g_{\text{eff}}} +\sqrt{-g_{\rm eff}}\mathcal{L}_\pi(g_{\rm eff},\pi)\,.
\end{equation}
with non-minimal derivative couplings in the matter Lagrangian $\mathcal{L}_\pi$ and the potential interactions in form of $\sqrt{g_{\text{eff}}}$. The composite effective metric has a very specific structure in terms of the $g$ and $f$ metrics \cite{deRham:2014naa}
\begin{equation}\label{effmetric}
g^\text{eff}_{\mu\nu}=\alpha^2 g_{\mu\nu} +2\alpha\beta
\,\mathcal{G}^\text{eff}_{\mu\nu} +\beta^2 f_{\mu\nu}\,,
\end{equation}
with $\alpha$ and $\beta$ being two arbitrary constants, and $\mathcal{G}^\text{eff}$ stands for
\begin{equation}\label{Geff}
\mathcal{G}^\text{eff}_{\mu\nu} =
g_{\mu\rho}X^\rho_\nu=f_{\mu\rho}Y^\rho_\nu\,,
\end{equation}
where we denoted the square root matrix by $X=\sqrt{g^{-1}f}$, and its inverse by $Y=\sqrt{f^{-1}g}$. The special property of this effective metric is the fact that its square root of the determinant corresponds to the allowed potentials
\begin{equation}\label{detgeff}
\sqrt{-g_\text{eff}}=\sqrt{-g} \,\det\bigl(\alpha 1 +\beta
X\bigr)=\sqrt{-f} \,\det\bigl(\beta 1 +\alpha Y\bigr)\,.
\end{equation}
We can express the square root of the determinant also in terms of the elementary symmetric polynomials $e_n(X)$ and $e_n(Y)$ as
\begin{equation}
\sqrt{-g_\text{eff}} = \sqrt{-g} \sum_{n=0}^4\alpha^{4-n}\beta^{n}
e_n(X) = \sqrt{-f} \sum_{n=0}^4\alpha^{n}\beta^{4-n} e_n(Y)\,,
\end{equation}
where the elementary symmetric polynomials have the following specific structure
{\allowdisplaybreaks 
\begin{subequations}\begin{eqnarray}\label{polynomials}
e_0(X)&=&1\,, \\ e_1(X)&=& \bigl[X\bigr]\,, \\ e_2(X)&=&
\frac{1}{2}\bigl(\bigl[X\bigr]^2-\bigl[X^2\bigr]\bigr)\,, \\ e_3(X)&=&
\frac{1}{6}\bigl(\bigl[X\bigr]^3-3\bigl[X\bigr]\bigl[X^2\bigr]
+2\bigl[X^3\bigr]\bigr)\,, \\ e_4(X)&=&
\frac{1}{24}\bigl(\bigl[X\bigr]^4-6\bigl[X\bigr]^2\bigl[X^2\bigr]
+3\bigl[X^2\bigr]^2 +
8\bigl[X\bigr]\bigl[X^3\bigr]-6\bigl[X^4\bigr]\bigr)\,.
\end{eqnarray}
\end{subequations}}

The Lagrangian for the non-minimal derivative couplings between the composite effective metric and the scalar field is given by
\begin{equation}\label{Proxy_geff}
\mathcal{L}_{\pi}=M_{\text{eff}} \(-a_1\pi R_{g_{\text{eff}}}-\frac{a_2}{\Lambda^3}\partial_\mu\pi\partial_\nu\pi G^{\mu\nu}_{g_{\text{eff}}}-\frac{a_3}{\Lambda^6}\partial_\mu\pi\partial_\nu\pi \Pi_{\alpha\beta} L^{\mu\alpha\nu\beta}_{g_{\text{eff}}}\)\,.
\end{equation}
In Lagrangian (\ref{Proxy_geff}), $G^{\mu\nu}_{g_{\text{eff}}}$ is the Einstein tensor of the effective metric and the tensor $L^{\mu\alpha\nu\beta}_{g_{\text{eff}}}$ corresponds to the dual Riemann tensor of the effective metric
\ba
L^{\mu\alpha\nu\beta}_{g_{\text{eff}}}&=&2R^{\mu\alpha\nu\beta}_{g_{\text{eff}}}+2(R^{\mu\beta}_{g_{\text{eff}}}g^{\nu\alpha}_{\text{eff}}+R^{\nu\alpha}g^{\mu\beta}_{\text{eff}}-R^{\mu\nu}_{g_{\text{eff}}}g^{\alpha\beta}_{\text{eff}}-R^{\alpha\beta}_{g_{\text{eff}}}g^{\mu\nu}_{\text{eff}}) \nonumber\\
&+&R_{g_{\text{eff}}}(g^{\mu\nu}_{\text{eff}}g^{\alpha\beta}_{\text{eff}}-g^{\mu\beta}_{\text{eff}}g^{\nu\alpha}_{\text{eff}})\,.
\ea
Coupling the matter field to this effective metric (instead of coupling it only to the space-time metric $g$) has important theoretical and phenomenological consequences. Particularly, it is a very interesting question to study the consequences of non-minimal derivative couplings to this effective metric. We could in principal consider all the Horndeski scalar-tensor interactions. However, in this work we would like to focus on a subclass of the Horndeski interactions given by (\ref{Proxy_geff}). This subclass is special in the sense that it is naturally generated by the covariantisation procedure of the decoupling limit of massive gravity discussed in \cite{deRham:2011by, Heisenberg:2014kea}.  Note, that we did not include an additional kinetic term for the $f$ metric in our Lagrangian (\ref{eq:TotalAction}) in contrast to the standard formulation of bigravity \cite{Hassan:2011zd}. The reasons for that will become clear in the next section, when we perform the Hamiltonian analysis of maximally symmetric space-times in the context of allowed kinetic terms in the presence of derivative non-minimal couplings to the matter fields.

As next, we shall compute the covariant equations of motion. First of all the Einstein equations are given by \cite{Hassan:2011vm, Schmidt-May:2014xla}\footnote{We will follow the same notation as in \cite{Blanchet:2015sra}.}
\begin{eqnarray}
\label{eq.:Einstein}
2\mpl^2 \,Y^{(\mu}_\rho G_g^{\nu)\rho} - m^2 M_\text{eff}^2
  \sum_{n=0}^3\alpha^{4-n}\beta^{n} \,Y^{(\mu}_\rho U_{(n)}^{\nu)\rho}&=&
  \alpha  M_\text{eff} \frac{\sqrt{-g_\text{eff}}}{\sqrt{-g}}\Bigl( \alpha
  Y^{(\mu}_\rho T_{g_{\text{eff}}}^{\nu)\rho}+\beta
  T_{g_{\text{eff}}}^{\mu\nu}\Bigr)\,, \nonumber\\
  - m^2 M_\text{eff}^2
  \sum_{n=0}^3\alpha^{n}\beta^{4-n} \,X^{(\mu}_\rho V_{(n)}^{\nu)\rho}
  &=&  \beta \frac{\sqrt{-g_\text{eff}}}{\sqrt{-f}}\Bigl( \beta
  X^{(\mu}_\rho T_{g_{\text{eff}}}^{\nu)\rho}+\alpha
  T_{g_\text{eff}}^{\mu\nu}\Bigr)\,,
  \end{eqnarray}
where the tensors $U_{(n)}^{\mu\nu}$ and
$V_{(n)}^{\mu\nu}$ stand for~\cite{Hassan:2011vm}
\begin{subequations}
\begin{align}
&U_{(n)} = \sum_{p=0}^n (-)^p e_{n-p}(X) X^p\,,\\ &V_{(n)} =
  \sum_{p=0}^n (-)^p e_{n-p}(Y) Y^p\,,
\end{align}
\end{subequations}
and the stress energy tensor is given by
\be
T^{g_{\text{eff}}}_{\mu\nu}=a_1T^{g_{\text{eff}}(1)}_{\mu\nu}-\frac{a_2}{\Lambda^3}T^{g_{\text{eff}}(2)}_{\mu\nu}-\frac{a_3}{\Lambda^6}T^{g_{\text{eff}}(3)}_{\mu\nu}.
\ee
Note that the properties of the Einstein and Riemann dual tensor ensure that the scalar field $\pi$ enters at most with two derivatives in the stress-energy tensor,
\ba
T^{g_{\text{eff}}(1)}_{\mu\nu}&=&\tau^{g_{\text{eff}}(1)}_{\mu\nu}+\pi G^{g_{\text{eff}}}_{\mu\nu} \label{eq:T1} \nonumber\\
T^{g_{\text{eff}}(2)}_{\mu\nu}&=&\tau^{g_{\text{eff}}(2)}_{\mu\nu}+\frac12L^{g_{\text{eff}}}_{\mu\alpha\nu\beta}\partial^\alpha\pi\partial^\beta\pi+\frac12G^{g_{\text{eff}}}_{\mu\nu}\mathcal{X} \label{eq:T2} \nonumber\\
T^{g_{\text{eff}}(3)}_{\mu\nu}&=&\tau^{g_{\text{eff}}(3)}_{\mu\nu}+\frac32L^{g_{\text{eff}}}_{\mu\alpha\nu\beta}\Pi^{\alpha\beta}_{g_{\text{eff}}}\mathcal{X} \label{eq:T3}
\ea
with $\mathcal{X}=(\partial\pi)^2$ and the tensors $\tau^{g_{\text{eff}}(1,2,3)}\mn$ encode the interactions of the scalar field 
\ba
\tau^{g_{\text{eff}}(1)}_{\mu\nu} &=&[\Pi_{g_{\text{eff}}}]g^{\text{eff}}_{\mu\nu}-\Pi^{g_{\text{eff}}}_{\mu\nu} \label{eq:X1} \\
\tau^{g_{\text{eff}}(2)}_{\mu\nu} &=& (\Pi^{g_{\text{eff}}}_{\mu\nu})^2-[\Pi_{g_{\text{eff}}}]\Pi_{\mu\nu}^{g_{\text{eff}}}-\frac12([\Pi_{g_{\text{eff}}}^2]-[\Pi_{g_{\text{eff}}}]^2)g^{\text{eff}}_{\mu\nu}\\
\tau^{g_{\text{eff}}(3)}_{\mu\nu} &=& 6(\Pi^{g_{\text{eff}}}_{\mu\nu})^3-6[\Pi_{^{g_{\text{eff}}}}](\Pi^{g_{\text{eff}}}_{\mu\nu})^2+3([\Pi_{g_{\text{eff}}}]^2-[\Pi_{g_{\text{eff}}}^2])\Pi^{g_{\text{eff}}}_{\mu\nu} \nonumber\\
&&-g_{\mu\nu}([\Pi_{g_{\text{eff}}}]^3-3[\Pi_{g_{\text{eff}}}^2][\Pi_{g_{\text{eff}}}]+2[\Pi_{g_{\text{eff}}}^3])\,,\label{eq:X3}
\ea
with the abbreviation $[\Pi_{g_{\text{eff}}}]=\nabla^{g_{\text{eff}}}_\mu \nabla_{g_{\text{eff}}}^\mu \pi$ for convenience. 
The equation of motion for the scalar field $\pi$ is given by
\ba
\mathcal{E}_\pi=
-a_1R_{g_{\text{eff}}}-\frac{2a_2}{\Lambda^3}G_{g_{\text{eff}}}^{\mu\nu}\Pi^{g_{\text{eff}}}_{\mu\nu}-\frac{2a_3}{\Lambda^6}L^{\mu\alpha\nu\beta}_{g_{\text{eff}}}(2\Pi_{\mu\nu}^{g_{\text{eff}}}\Pi_{\alpha\beta}^{g_{\text{eff}}}+R^{(g_{\text{eff}})\gamma}_{\;\;\beta\alpha\nu}\partial_\gamma\pi\partial_\mu\pi)=0\,,\label{eq:piJ}
\ea
where we made use of
\ba
\nabla^{g_{\text{eff}}}_\nu \nabla^{g_{\text{eff}}}_\alpha \nabla^{g_{\text{eff}}}_\beta\pi L_{g_{\text{eff}}}^{\mu\alpha\nu\beta}&=&\frac12R^{(g_{\text{eff}})\gamma}_{\;\;\beta\alpha\nu}\partial_\gamma\pi L^{\mu\alpha\nu\beta}_{g_{\text{eff}}}=-\frac18 \partial^\mu\pi \mathcal L^{g_{\text{eff}}}_{\rm GB}
\ea
with the Gauss-Bonnet term of the effective metric $\mathcal L^{g_{\text{eff}}}_{\rm GB}=R^2_{g_{\text{eff}}}+\left(R^{g_{\text{eff}}}_{\mu\alpha\nu\beta}\right)^2-4\left(R^{g_{\text{eff}}}_{\mu\nu}\right)^2$. In the following we would like to study the consequences of the presence of the derivative non-minimal couplings in the action (\ref{eq:TotalAction}). 

\section{Hamiltonian analysis}
In this section we would like to perform the Hamiltonian analysis in the mini-superspace and discuss different scenarios. For this purpose we shall concentrate on the first non-minimal coupling term $\pi R_{g_{\text{eff}}}$ in (\ref{Proxy_geff}). The conclusions we make here for this first interaction also apply to the remaining interactions. We will consider three different scenarios concerning the dynamics of the $f$ metric appearing in $g_{\text{eff}}$.

{\bf 1. Non-dynamical $f$ metric:} The first scenario corresponds to the case where the $f$ metric is not dynamical and we choose it to be Minkowski. Thus, we have
\ba
\d s_g^2=- N(t)^2\d t^2 +a(t)^2 \d\vec{x}{}^2\,, \quad \quad{\rm and}\quad \quad \d s_f^2=- \d t^2 + \d\vec{x}{}^2 \,.
\ea
As we mentioned above, we will focus on the sub Lagrangian 
\begin{equation}\label{case1}
\mathcal{L}^{(1)}= \mpl^2 \sqrt{-g} R_g+ m^2M_{\text{eff}}^2\sqrt{g_{\text{eff}}}- M_{\text{eff}}a_1 \sqrt{g_{\text{eff}}} \pi R_{g_{\text{eff}}}\,.
\end{equation}
but all the statements that we make here apply also to the remaining non-minimal couplings of our Lagrangian (\ref{Proxy_geff}). The conjugate momenta associated with the $\pi$ field and the scale factor are
\begin{equation}
p_{\pi} =\frac{6M_{\text{eff}}a_1\alpha a_{\text{eff}}^2\dot{a}}{N_{\text{eff}}} \,,
\quad
p_{g} =-\frac{12\mpl^2a\dot{a}}{N} +\frac{6a_1M_{\text{eff}}\alpha a_{\text{eff}}(2\alpha \pi \dot{a}+a_{\text{eff}}\dot{\pi})}{N_{\text{eff}}}\,, 
\end{equation}
with the effective lapse function $ N_{\text{eff}}(t)=\alpha N(t)+\beta$ and the scale factor $a_{\text{eff}}(t) = \alpha a(t) + \beta$.
After performing the Legendre transformation we obtain that the Hamiltonian has a piece that is non-linear in the lapse
\begin{equation}
\mathcal{H}^{(1)} \supset \frac{\mpl^2 p_{\pi}^2\beta^2 a}{6a_1^2 M_{\text{eff}}^2\alpha^2 a_{\text{eff}}^4N} \,.
\end{equation}
This is a clear indication that we have lost the constraint equation that removes the Boulware-Deser ghost. The presence of the derivative non-minimal coupling with non-dynamical $f$ metric yields an unavoidable ghost degree of freedom. This is in complete agreement with the findings in \cite{Gao:2014xaa}, where they considered couplings in form of Horndeski scalar-tensor interactions living in the composite effective metric. In the case they studied, the $f$ metric was not dynamical and their perturbation analysis about a FLRW background indicated the presence of the ghost degree of freedom. Here we have seen that the Hamiltonian is always non-linear in the lapse in the presence of this type of derivative non-minimal couplings. However, as we shall see as next, we gain more freedom in making the $f$ metric dynamical as well and the Hamiltonian becomes automatically linear in the lapses if we do not add an additional kinetic term for the $f$ metric. Actually, this same conclusion applies to the pure kinetic Lagrangian without the non-minimal derivative coupling. If one for instance takes the kinetic terms $M_g^2\sqrt{-g} R_g + M_\text{eff}^2\sqrt{-g_\text{eff}} R_\text{eff}$ with non-dynamical $f$ metric, then this restricts the phase space so much that the Hamiltonian becomes non-linear in the lapse. If one allows the $f$ metric to be dynamical but without including its kinetic term, this enriches the phase space and the Hamiltonian has only then the chance to be linear in the lapses \cite{Blanchet:2015sra}.

{\bf 2. Dynamical $f$ metric without additional kinetic term:} In this scenario we assume that the $f$ metric also carries dynamics in $ R_{g_{\text{eff}}}$ but without including an additional kinetic term for the $f$ metric. This is a crucial distinction between this scenario and the next one. So our starting point is again the sub Lagrangian (\ref{case1}) but this time with
\ba
\d s_g^2=- N(t)^2\d t^2 +a(t)^2 \d\vec{x}{}^2\,, \quad \quad{\rm and}\quad \quad \d s_f^2=- n_f(t)^2\d t^2 +a_f(t)^2 \d\vec{x}{}^2 \,.
\ea
In this case the conjugate momenta associated with the $\pi$ field and the two scale factors become

\begin{eqnarray}
p_{\pi} =\frac{6a_1M_{\text{eff}} a_{\text{eff}}^2\dot{a}_{\text{eff}}}{N_{\text{eff}}} \,,
\quad
p_{f} =\beta \mathcal{Q} \,, \quad
p_{g} =-\frac{12\mpl^2a\dot{a}}{N}+\alpha \mathcal{Q}
\end{eqnarray}
where we have defined 
\begin{equation}
\mathcal{Q}=\frac{6a_1 M_{\text{eff}} a_{\text{eff}}(2\pi\dot{a}_{\text{eff}}+a_{\text{eff}}\dot{\pi}) }{N_{\text{eff}}}
\end{equation}
This time the Hamiltonian 
\begin{equation}
\mathcal{H}^{(2)} = \dot{a}p_g + \dot{a}_f \pi_f+\dot{\pi} p_{\pi}-\mathcal{L}^{(1)} \,,
\end{equation}
is automatically linear in the lapses. Thus, giving dynamics to the $f$ metric rescues the derivative non-minimal couplings. The linearity of the Hamiltonian in the lapses ensures the propagation of the constraint that on the other hand removes the Boulware-Deser ghost. 

{\bf 3. Dynamical $f$ metric with an additional kinetic term:} In this third scenario we shall consider the dynamical case as in the second scenario but with an additional kinetic term for the $f$ metric. This has severe consequences. Our sub Lagrangian this time consists of
\begin{equation}\label{case3}
\mathcal{L}^{(3)}= \mpl^2 \sqrt{-g} R_g +M_f^2 \sqrt{-f} R_f+ m^2M_{\text{eff}}^2\sqrt{g_{\text{eff}}} - a_1M_{\text{eff}} \sqrt{g_{\text{eff}}} \pi R_{g_{\text{eff}}}\,.
\end{equation}
In this scenario the conjugate momenta for the scale factor of the $f$ metric obtains an additional term while the others remain the same

\begin{eqnarray}
p_{\pi} =\frac{6a_1M_{\text{eff}} a_{\text{eff}}^2\dot{a}_{\text{eff}}}{N_{\text{eff}}} \,,
\quad
p_{f} =-\frac{12M_f^2a_f\dot{a}_f}{n_f}+\beta \mathcal{Q} \,, \quad
p_{g} =-\frac{12\mpl^2a\dot{a}}{N}+\alpha \mathcal{Q}
\end{eqnarray}
After expressing the time derivatives of the fields in terms of their conjugate momenta, one obtains that the Hamiltonian is always non-linear in the lapses independently of what one choses for the coefficient in front of the non-minimal coupling. This non-linear contributions are proportional to $M_f^2 \mpl^2 \alpha^2 \beta^2$. In order for the Hamiltonian to remain linear in the lapses in the presence of the derivative non-minimal couplings, one has to restrict himself to the second scenario. It is mandatory to have both metrics dynamical but without the inclusion of the second kinetic term. This coincides with the findings of \cite{Blanchet:2015sra} where it was shown that one can not have three kinetic terms at the same time. Here we saw also that exactly the same happens in the case of derivative non-minimal couplings to matter fields. We have checked explicitly that these conclusions also apply to the remaining derivative couplings $\sqrt{-g_{\text{eff}}}\partial_\mu\pi\partial_\nu\pi G^{\mu\nu}_{g_{\text{eff}}}$ and $\sqrt{-g_{\text{eff}}}\partial_\mu\pi\partial_\nu\pi \Pi_{\alpha\beta} L^{\mu\alpha\nu\beta}_{g_{\text{eff}}}$. 

Summarising, among the three different scenarios that we have considered here, only the second option yields a consistent way of coupling. If one insists on non-minimal derivative couplings of the composite effective metric to the matter fields, then this excludes the more restrictive case of fixed $f$ metric. It would be very interesting to perform the same cosmological perturbation analysis as in \cite{Gao:2014xaa} but keeping the second metric dynamical as well without the inclusion of its kinetic term.

\section{Decoupling Limit}

In this section we will pay special attention to the non-minimal couplings between the scalar field $\pi$ and the effective metric in the decoupling limit. First we restore the broken diffeomorphism by introducing the Stueckelberg fields in the $f$ metric
\begin{equation}
f_{\mu\nu} \to \tilde f_{\mu\nu} = f_{ab}\partial_\mu \mathcal{S}^a
\partial_\nu \mathcal{S}^b\,.
\end{equation}
These Stueckelberg fields $\mathcal{S}^a$ can be further decomposed into the
helicity-0 mode $\phi$ and helicity-1 mode $A^a$,
\begin{equation}
\mathcal{S}^a=x^a-\frac{m A^a}{\Lambda^3_3} -\frac{f^{ab}\partial_b
  \phi}{\Lambda^3_3}\,.
\end{equation}
Now we can take the decoupling limit consisting of sending $\mpl \to
\infty$, while keeping the scale $\Lambda^3_3=\mpl m^2$ fixed. Let us mention that we will neglect the contribution of the
helicity-1 field and keep track of the contributions of the
helicity-0 mode $\phi$ only. We would like to see whether or not the equation of motion for the helicity-0 mode
contains higher derivative terms due to the derivative non-minimal couplings. For this, we will concentrate on the $\mathcal{L}_\pi$ part of the action (\ref{eq:TotalAction}) and we will further assume $g_{\mu\nu}=\eta_{\mu\nu}$ in this section and
\begin{equation}
\label{eq:fDL}
f_{\mu\nu} = \eta_{\mu\nu} \to \tilde f_{\mu\nu} =
\left(\eta_{\mu\nu}-\Phi_{\mu\nu}\right)^2\,,
\end{equation}
where $\Phi_{\mu\nu}$ stands for $\Phi_{\mu\nu} \equiv \partial_\mu
\partial_\nu \phi/\Lambda^3_3$. In this case the effective metric in the decoupling limit
becomes
\begin{equation}
g_{\mu\nu}^{\text{eff}} \to \tilde{g}_{\mu\nu}^{\text{eff}} = \bigl[
  (\alpha+\beta)\eta_{\mu\nu}-\beta \Phi_{\mu\nu} \bigr]^2\,.
\end{equation}
The contribution of the derivative non-minimal couplings $\mathcal{L}_\pi$ to the equation of motion for the helicity-0 mode is given by
\begin{eqnarray}
\frac{\delta \mathcal{L}_{\pi}}{\delta \phi}&=&  \frac{\partial_\mu \partial_\nu}{\Lambda^3}\left( \frac{\delta \mathcal{L}_{\pi}}{\delta \tilde{g}^{\text{eff}}_{\rho\sigma}} \frac{\delta \tilde{g}^{\text{eff}}_{\rho\sigma}}{\delta \Phi_{\mu\nu}} \right) \nonumber\\
&=& -\frac{\partial_\mu \partial_\nu}{\Lambda^3} \Big( \beta  \sqrt{-\tilde{g}_{\text{eff}}}T^{\mu\rho}_{\tilde{g}_\text{eff}}((\alpha+\beta)\delta^\nu_\rho-\beta\Phi^\nu_\rho)  \Big)\nonumber\\
&=&-\frac{
  \partial_\nu}{\Lambda^3_3} \left\{\beta \partial_\mu(
\sqrt{-\tilde{g}_{\text{eff}}}T^{\mu\rho}_{\text{eff}})
\bigl[(\alpha+\beta)\delta^\nu_\rho-\beta\Phi^\nu_\rho\bigr]
-\beta^2\sqrt{-\tilde{g}_{\text{eff}}}\,T^{\mu\rho}_{\tilde{g}_{\text{eff}}}\,
\partial_\mu \Phi^\nu_\rho\right\} \,.
\end{eqnarray}
The stress energy tensor is tranverse on-shell 
\begin{equation}
\nabla^{\tilde{g}_{\text{eff}}}_\mu T_{\tilde{g}_{\text{eff}}}^{\mu\rho}= \frac{1}{\sqrt{-\tilde{g}_{\text{eff}}}} \partial_\mu\left(
\sqrt{-\tilde{g}_{\text{eff}}}T^{\mu\rho}_{\tilde{g}_{\text{eff}}}\right)
+
\Gamma_{\mu\sigma}^{\rho\,\tilde{g}_{\text{eff}}}\,T^{\mu\sigma}_{\tilde{g}_{\text{eff}}} =\partial^\rho \pi
\mathcal{E}_\pi\,,
\end{equation}
where $\mathcal{E}_\pi$ denotes the equation of motion of the matter field $\pi$ given by (\ref{eq:piJ}). Using this we can write the equation of motion for the helicity-0 mode as
\begin{align}
\frac{\delta \mathcal{L}_{\pi}}{\delta \phi} = -\frac{
  \partial_\nu}{\Lambda^3_3} & \left\{  \beta
\left(\sqrt{-\tilde{g}_{\text{eff}}} \,\partial^\rho \pi
\mathcal{E}_\pi
-\sqrt{-\tilde{g}_{\text{eff}}}
\,\Gamma_{\mu\sigma}^{\rho\,\tilde{g}_{\text{eff}}}\,T^{\mu\sigma}_{\tilde{g}_{\text{eff}}}
\right) \bigl[(\alpha+\beta)\delta^\nu_\rho-\beta\Phi^\nu_\rho\bigr] \right. \nonumber\\
&\left. -\beta^2\sqrt{-\tilde{g}_{\text{eff}}}
\,T^{\mu\rho}_{\tilde{g}_{\text{eff}}}\partial_\mu
\Phi^\nu_\rho\right\} \,. \nonumber
\end{align}
This can be similarly written as
\begin{eqnarray}
\frac{\delta \mathcal{L}_{\pi}}{\delta \phi}&=& -\frac{
  \partial_\nu}{\Lambda^3_3} \left\{
 \beta \sqrt{-\tilde{g}_{\text{eff}}}
\partial^\rho \pi
\mathcal{E}_\pi
\bigl[(\alpha+\beta)\delta^\nu_\rho-\beta\Phi^\nu_\rho\bigr]
-\beta^2\sqrt{-\tilde{g}_{\text{eff}}}T^{\mu\sigma}_{\tilde{g}_{\text{eff}}}
\mathcal{R}_{\mu\sigma}^{\nu}\right\}\,,
\end{eqnarray}
where 
\begin{equation}
 \mathcal{R}^{\nu}_{\mu\sigma}
=
\Gamma^{\rho\,\tilde{g}_{\text{eff}}}_{\mu\sigma}\bigl[(\alpha+\beta)
  \delta^\nu_\rho-\Phi^\nu_\rho\bigr]+\beta^2\partial^\nu\Phi_{\mu\sigma}\,.
\end{equation}
We can express the Christoffel symbols in terms of the helicity-0 mode as
\begin{equation}
\Gamma^{\rho\,\tilde{g}_{\text{eff}}}_{\mu\sigma}
=-\tilde{g}_{\text{eff}}^{\rho\kappa}\bigl[(\alpha+\beta)\delta^\nu_\kappa
  -\beta\Pi^\nu_\kappa\bigr]\partial_\nu \Pi_{\mu\sigma}\,,
\end{equation}
which together with the relation 
\begin{equation}
\bigl[(\alpha+\beta)\delta^\mu_\rho-\beta\Phi^\mu_\rho\bigr]
\tilde{g}_{\text{eff}}^{\rho\sigma}\bigl[(\alpha+\beta)\delta^\nu_\sigma
  -\beta\Phi^\nu_\sigma\bigr]=\eta^{\mu\nu}\,,
\end{equation}
implies that $\mathcal{R}^{\nu}_{\mu\sigma}=0$ and we are left with
\begin{eqnarray}
\frac{\delta \mathcal{L}_{\pi}}{\delta \phi}&=& -\frac{
  \partial_\nu}{\Lambda^3_3} \left\{
 \beta \sqrt{-\tilde{g}_{\text{eff}}}
\partial^\rho \pi
\mathcal{E}_\pi
\bigl[(\alpha+\beta)\delta^\nu_\rho-\beta\Phi^\nu_\rho\bigr]
\right\}\,.
\end{eqnarray}
We see immediately that the equation of motion for the helicity-0 mode in the decoupling limit remains second order in derivatives even in the presence of the non-minimal couplings between the scalar field $\pi$ and the effective metric, after taking into account the equation of motion for the matter field.

\section{Cosmological background}\label{sec:SA}
In this section we would like to compute the field equations of the theory (\ref{eq:TotalAction}) on cosmological backgrounds. In a future work we shall push forward these preliminary studies and investigate the cosmological perturbations. Let us consider a FLRW background for the metric $g$ and $f$,
\ba
\d s_g^2=- N(t)^2\d t^2 +a(t)^2 \d\vec{x}{}^2\,, \quad \quad{\rm and}\quad \quad \d s_f^2=- n_f(t)^2\d t^2 +a_f(t)^2 \d\vec{x}{}^2 \,.
\ea
with $a(t)$ being the scale factor. The effective metric then takes the form
\ba
\d s_{g_{\text{eff}}}^2=-N_{\text{eff}}(t)^2  \d t^2 + a_{\text{eff}}(t)^2  \d\vec{x}{}^2\,, 
\ea
with the effective lapse function and the scale factor as
\ba
 N_{\text{eff}}(t)=\alpha N(t)+\beta n_f(t)\quad{\rm and}\quad  a_{\text{eff}}(t) = \alpha a(t) + \beta a_f(t) \,.
\ea
We can further define the Hubble parameters $H=\dot{a}/a$, $H_f=\dot{a}_f/a_f$ and $H_{\text{eff}}=\dot{a}_{\text{eff}}/a_{\text{eff}}$.
Let us first compute the resulting effective energy density and pressure for the field $\pi$ 
\ba
\rho^\pi&=& 6a_1\left(H_{\text{eff}}^2\pi+\frac{H_{\text{eff}}}{N_{\text{eff}}}\dot\pi\right)-\frac{9a_2}{\Lambda^3}\frac{H_{\text{eff}}^2}{N_{\text{eff}}^2}\dot\pi^2-\frac{30a_3}{\Lambda^6}\frac{H_{\text{eff}}^3}{N_{\text{eff}}^3}\dot\pi^3 \label{eq:effdens}\\
P^\pi&=&2\left[\frac{6a_3}{\Lambda^6}H_{\text{eff}}\dot\pi^2(\frac{\dot\pi}{N_{\text{eff}}^4}(\dot H_{\text{eff}}N_{\text{eff}}+H_{\text{eff}}^2N_{\text{eff}}^2-\frac{3}{2} H_{\text{eff}}\dot{N}_{\text{eff}})+\frac32\frac{H_{\text{eff}}}{N_{\text{eff}}^3}\ddot\pi) \right. \nonumber\\
&&+\frac{a_2}{2\Lambda^3}\dot\pi(\frac{\dot\pi}{N_{\text{eff}}^3}(3H_{\text{eff}}^2N_{\text{eff}}^2+2\dot H_{\text{eff}}N_{\text{eff}}-4 H_{\text{eff}}\dot{N}_{\text{eff}})+4\frac{H_{\text{eff}}}{N_{\text{eff}}^2}\ddot\pi) \nonumber\\
&&\left. -a_1(\pi (3H_{\text{eff}}^2N_{\text{eff}}+2\dot H_{\text{eff}})+\dot\pi(2H_{\text{eff}}- \frac{\dot{N}_{\text{eff}}}{N_{\text{eff}}^2})+\frac{\ddot\pi}{N_{\text{eff}}})\right]\,,\label{eq:effpress}
\ea
in terms of which the Einstein equations for the $g$ metric can be expressed as
\ba
6\mpl^2 H^2+m_2 M_{\text{eff}}^2\alpha\frac{a_{\text{eff}}^3}{a^3}&=&\alpha M_{\text{eff}} \frac{a_{\text{eff}}^3}{a^3}\rho_\pi \nonumber\\
-2 \mpl^2 (2\dot{H}+3H^2N)+m^2M_{\text{eff}}^2\alpha N_{\text{eff}} \frac{a_{\text{eff}}^2}{a^2}&=&\alpha M_{\text{eff}} \frac{a_{\text{eff}}^2}{a^2}P_\pi \,.
\ea
Similarly the Einstein equations for the $f$ metric are given by
\ba
m_2 M_{\text{eff}}^2\beta\frac{a_{\text{eff}}^3}{a_{f}^3}&=&\beta M_{\text{eff}} \frac{a_{\text{eff}}^3}{a_f^3}\rho_\pi \nonumber\\
m^2M_{\text{eff}}^2\beta N_{\text{eff}} \frac{a_{\text{eff}}^2}{a^2}&=&\beta M_{\text{eff}} \frac{a_{\text{eff}}^2}{a_f^2}P_\pi
\ea
On the other hand the equation of motion for the field $\pi$ on the FLRW background yields
\ba
&&\frac{6a_2}{\Lambda^3}\(\frac{H_{\text{eff}}\dot\pi}{N_{\text{eff}}^2}(3H_{\text{eff}}^2N_{\text{eff}}^2+2\dot H_{\text{eff}}N_{\text{eff}}- H_{\text{eff}} \dot{N}_{\text{eff}})+\frac{H_{\text{eff}}^2}{N_{\text{eff}}}\ddot\pi\) \nonumber\\
&&+\frac{18a_3}{\Lambda^6}\(\frac{H_{\text{eff}}^2 \dot\pi^2}{N_{\text{eff}}^3}\left( 3N_{\text{eff}}(H_{\text{eff}}^2N_{\text{eff}}+\dot H_{\text{eff}})-2H_{\text{eff}}\dot{N}_{\text{eff}} \right)+2\frac{H_{\text{eff}}^3}{N_{\text{eff}}^2}\dot\pi\ddot\pi\)=a_1\bar{R}_{\text{eff}}.
\ea
where we have defined $\bar{R}_{\text{eff}}=6(\dot{H}_{\text{eff}}+2H_{\text{eff}}^2N_{\text{eff}} )$. Here we have quoted the background equations of motion for our Lagrangian. They are quite involved but in principle solvable. The phase map is very reach due to the numerous degrees of freedom. In a follow-up work we would like to study the dynamical system of cosmological solutions by means of a phase map analysis and explore the critical points of the cosmological equations and compare with numerical solutions. It would be also interesting to see if a bouncing solution at early times could be possible. Furthermore, we would like to investigate the cosmological perturbations on top of this general FLRW background and study the stability conditions.

\section{Discussion}\label{sec:conclusion}
In this work we have studied the theoretical implication of non-minimal derivative couplings of the effective composite metric, proposed in \cite{deRham:2014naa}, to matter fields. Instead of considering the entire class of Horndeski scalar-tensor interactions, we focused on a very specific subclass of non-minimal derivative interactions. The motivation for this comes from the fact, that these interactions are naturally generated by the covariantisation procedure of the decoupling limit of massive gravity. There, the specific interactions between the helicity-0 and helicity-2 modes generate these specific couplings between a scalar field and terms that depend on the Riemann tensor and Ricci tensor..etc. These curvature terms live in the usual space-time metric. Here we have promoted these derivate interactions to those living in the composite effective metric. We first computed the covariant equations of motion. In the same way as in the original proxy theory, the Einstein and the double dual Riemann tensor are divergenceless, which then only generates second order equations of motion but in terms of the composite effective metric. As next we performed a detailed Hamiltonian analysis of maximally symmetric space-times and considered three different scenarios. In the first scenario we assumed that the second metric is fixed and we saw immediately that the non-minimal derivative coupling makes the Hamiltonian to become non-linear in the lapse. In the second scenario we allowed the second metric to be dynamical as well but without including an additional kinetic term. In this case the Hamiltonian becomes automatically linear in the lapses. Finally, in the third scenario we included an additional kinetic term for the second metric and the consequence of this case was a Hamiltonian non-linear in the lapses. Thus, we concluded that the only consistent way of non-minimal derivative couplings is through the second scenario with the dynamical $f$ metric without an additional kinetic term. This is in a very close analogy to the studies in \cite{Blanchet:2015sra} in the context of allowed kinetic terms. Making the second metric dynamical enriches the phase space and gives more freedom. However, this has to be done without the inclusion of a second kinetic term. The conclusions of the first scenario with fixed $f$ metric coincides with what was obtained in \cite{Gao:2014xaa} for the Horndeski class. It would be very interesting to push forward their analysis of cosmological perturbations to the case of the second scenario with the dynamical $f$ metric with no further kinetic term. We shall investigate this further in a future work. Furthermore, we investigated the non-minimal derivative couplings in the decoupling limit and saw that the equation of motion for the helicity-0 mode remains second order in derivatives after taking into account the equation of motion for the scalar field. Moreover, we computed the field equation on FLRW backgrounds and we shall investigate the numerical solutions together with the cosmological perturbations in a future work.

\section*{Acknowledgments}

L.H. wishes to acknowledge the African Institute for Mathematical Sciences in Muizenberg, South Africa, for hospitality and support at the latest stage of this work.

\bibliographystyle{JHEPmodplain}
\bibliography{deriv_geffs.bib}

\end{document}